\documentclass[prd,preprint,nofootinbib]{revtex4}
\usepackage{amssymb}
\usepackage{epsfig}
\begin{document}
\title{ 
 Thermal leptogenesis in brane world cosmology
}

\author{Nobuchika Okada}
 \email{okadan@post.kek.jp}
 \affiliation{
 Theory Division, KEK, Oho 1-1, Tsukuba, Ibaraki 305-0801, Japan \\ 
 Department of Particle and Nuclear Physics, 
 The Graduate University \\ 
 for Advanced Studies (Sokendai), 
 Oho 1-1, Tsukuba, Ibaraki 305-0801, Japan 
 }

\author{Osamu Seto}
 \email{O.Seto@sussex.ac.uk}
 \affiliation{
 Department of Physics and Astronomy, University of Sussex, 
 Brighton BN1 9QJ, United Kingdom
}

%
\begin{abstract}
The thermal leptogenesis in brane world cosmology is studied.
In brane world cosmology, the expansion law is modified 
 from the four-dimensional standard cosmological one 
 at high temperature regime in the early universe. 
As a result, the well-known upper bound on 
 the lightest light neutrino mass 
 induced by the condition for 
 the out-of-equilibrium decay of the lightest heavy neutrino, 
 $\tilde{m}_1 \lesssim 10^{-3}$ eV, 
 can be moderated to be 
 $\tilde{m}_1 \lesssim 10^{-3} \; \mbox{eV} \times (M_1/T_t)^2 $ 
 in the case of $T_t \leq M_1$ 
 with the lightest heavy neutrino mass ($M_1$) and 
 the ``transition temperature'' ($T_t$), 
 at which the modified expansion law in brane world cosmology 
 is smoothly connecting with the standard one. 
This implies that the degenerate mass spectrum of the light neutrinos 
 can be consistent with the thermal leptogenesis scenario. 
Furthermore, as recently pointed out, 
 the gravitino problem in supersymmetric case can be solved 
 if the transition temperature is low enough 
 $T_t \lesssim 10^{6-7}$ GeV. 
Therefore, even in the supersymmetric case, 
 thermal leptogenesis scenario 
 can be successfully realized in brane world cosmology. 
\end{abstract}

\preprint{KEK-TH-1019} 

\vspace*{3cm}
\maketitle


\section{Introduction}

The origin of the cosmological baryon asymmetry is one of 
 the prime open questions in particle physics as well as in cosmology.
The asymmetry must have been generated during the evolution of the universe.
In fact, such a generation is possible if three conditions,
 \textit{i}) the existence of baryon number violating interactions, 
 \textit{ii}) C and CP violations and 
 \textit{iii}) the departure from thermal equilibrium, are 
 satisfied \cite{SakharovCond}.

Among various mechanisms of baryogenesis, leptogenesis \cite{FukugitaYanagida}
 is attractive because of its simplicity and 
 the connection to neutrino physics. 
Particularly, the simplest scenario, namely thermal leptogenesis, requires
 nothing but the thermal excitation of heavy Majorana neutrinos which generate 
 tiny neutrino masses via the seesaw mechanism \cite{seesaw} and 
 provides several implications for the light neutrino mass spectrum 
 \cite{Buchmulleretal}.
In leptogenesis, the first condition is satisfied by the Majorana nature of
 heavy neutrinos and the sphaleron effect in the standard model (SM) 
 at the high temperature \cite{KRS}, 
 while the second condition is provided by their CP violating decay. 
The departure from thermal equilibrium is provided by the expansion of
 the universe. 

The out-of-equilibrium decay is realized if the decay rate is smaller 
 than the expansion rate of the universe,
\begin{equation} 
\Gamma_{N_1} < \left.H\right|_{T=M_1} , 
 \label{OutOfEquilibDecay}
\end{equation}
 where $\Gamma_{N_1}$ and $M_1$ are the decay rate 
 and the mass of the lightest heavy neutrino, respectively, 
 and $H$ denotes the Hubble parameter. 
Note that the expansion law is governed by the gravitational theory.
Therefore, if general relativity is replaced by another 
 theory at a high energy scale, 
 the universe would undergo non-standard expansion.
One of such examples is the brane world cosmology \cite{braneworld}. 
The following discussion is based on 
 the so-called ``RS II'' model first proposed 
 by Randall and Sundrum \cite{RS}. 
In the model, the Friedmann equation for a spatially flat spacetime 
 is given by
\begin{equation}
H^2 = \frac{8\pi G_N}{3}\rho\left(1+\frac{\rho}{\rho_0}\right) 
\label{BraneFriedmannEq}
\end{equation}
 where 
\begin{equation}
\rho = \frac{\pi^2}{30}g_*T^4
\end{equation}
 is the energy density of the radiation with $g_*$ being 
 the effective degrees of freedom of relativistic particles, 
\begin{eqnarray}
\rho_0 = 96 \pi G_N M_5^6, \label{M5}
\end{eqnarray}
 with $M_5$ being the five dimensional Planck mass
\footnote{
We define $M_5$ as the ``reduced'' five dimensional Planck mass. 
In some papers, the normal five dimensional Planck mass is used. 
The reduced five dimensional Planck mass is defined 
 as $M_5/(8 \pi)^{1/3}$ by using the normal one ($M_5$). 
}, 
and the four dimensional 
 cosmological constant has been tuned to be zero. 
Here we have omitted the term so-called ``dark radiation'', 
 since this term is severely constrained by 
 big bang nucleosynthesis \cite{ichiki} and
 does not affect the results in this paper.
The second term proportional to $\rho^2$ 
 is a new ingredient in the brane world cosmology 
 and leads to a non-standard expansion law. 

Note that according to Eq.~(\ref{BraneFriedmannEq}) 
 the evolution of the early universe can be divided into two eras. 
At the era where $\rho/\rho_0 \gg 1$ the second term dominates 
 and the expansion law is non-standard (brane world cosmology era), 
 while at the era $\rho/\rho_0 \ll 1$ the first term dominates 
 and the expansion of the universe 
 obeys the standard expansion law (standard cosmology era). 
In the following, we call a temperature defined as 
 $\rho(T_t)/\rho_0=1$ ``transition temperature'', 
 at which the evolution of the early universe 
 changes from the brane world cosmology era 
 into the standard one.
The transition temperature $T_t$ is determined as
\begin{equation}
T_t \simeq 
 1.6 \times 10^7 \left(\frac{100}{g_*}\right)^{1/4}
 \left(\frac{M_5}{10^{11} {\rm GeV}}\right)^{3/2} {\rm GeV},
\end{equation}
 once $M_5$ is given.
Using the transition temperature, 
 we rewrite Eq.~(\ref{BraneFriedmannEq}) into the form,  
\begin{eqnarray}
H^2 = \frac{8\pi G_N}{3}\rho 
 \left[ 1+ \left(\frac{T}{T_t} \right)^4 \right]  
 = H_{st}^2 \left[1+ \left(\frac{T}{T_t} \right)^4 \right] , 
\label{BraneFriedmannEqII}
\end{eqnarray}
where $H_{st}$ is the Hubble parameter in the standard cosmology. 

This modification of the expansion law at a high temperature 
 ($T > T_t$) leads some drastic changes 
 for several cosmological issues. 
In fact, some interesting consequences in the brane world cosmology 
 such as the enhancement of the dark matter relic density \cite{OS-NOS}
 and the suppression of the overproduction of gravitino 
 \cite{OSGravitino} have been pointed out. 
In this paper, we investigate how the modified expansion law 
 in the brane world cosmology affects the thermal leptogenesis. 
Clearly, if $T_t < M_1$, 
 we can expect some effects 
 according to the condition of Eq.~(\ref{OutOfEquilibDecay}).

\section{A brief overview of thermal leptogenesis}

In the seesaw model, the smallness of the neutrino masses can be 
 naturally explained by the small mixings 
 between left-handed neutrinos and 
 heavy right-handed Majorana neutrinos $N_i$. 
The basic part of the Lagrangian in the SM 
 with right-handed neutrinos is described as 
\begin{eqnarray}
{\cal L}_{N}=-h_{ij} 
\overline{l_{L,i}} H N_j 
-\frac{1}{2} \sum_{i} M_i \overline{ N^C_i} N_i + h.c. , 
\label{Yukawa}
\end{eqnarray} 
where $i,j=1,2,3$ denote the generation indices, 
 $h$ is the Yukawa coupling, 
 $l_L$ and $H$ are the lepton and the Higgs doublets, 
 respectively, and 
 $M_i$ is the lepton-number-violating mass term  
 of the right-handed neutrino $N_i$ 
 (we are working on the basis of 
 the right-handed neutrino mass eigenstates). 
In this paper, we assume the hierarchical mass spectrum 
 for the heavy neutrinos, $M_1 \ll M_2, M_3$, 
 for simplicity as in many literature. 

In the case of the hierarchical mass spectrum for the heavy neutrinos, 
 the lepton asymmetry in the universe is generated 
 dominantly by CP-violating out-of-equilibrium decay of 
 the lightest heavy neutrino, $N_1 \rightarrow l_L H^*$ 
 and $ N_1 \rightarrow \overline{l_L} H $. 
The leading contribution is given by the interference 
 between the tree level and the one-loop level decay amplitudes, 
 and the CP-violating parameter is described as \cite{FandG, SUSYFandG} 
\begin{eqnarray}
\varepsilon &\equiv& 
\frac{\Gamma(N_1\rightarrow H+\bar{l}_j)-\Gamma(N_1\rightarrow H^*+l_j)}
{\Gamma(N_1\rightarrow H+\bar{l}_j)+\Gamma(N_1\rightarrow H^*+l_j)} \\
&\simeq& \frac{1}{8\pi}\frac{1}{(h h^{\dagger})_{11}}\sum_{i=2,3}
 \textrm{Im} (hh^{\dagger})^2_{1i}
 \left[f\left(\frac{M_i^2}{M_1^2}\right)+ 
 2 g\left(\frac{M_i^2}{M_1^2}\right)\right] . 
\end{eqnarray}
Here $f(x)$ and $g(x)$ correspond to 
 the vertex and the wave function corrections, 
\begin{eqnarray}
f(x)&\equiv& \sqrt{x} \left[  
1-(1+x)\mbox{ln} \left(\frac{1+x}{x} \right) \right] ,
 \nonumber \\
g(x)&\equiv& \frac{\sqrt{x}}{2(1-x)} ,  
\end{eqnarray}
 respectively. 
These functions are slightly modified 
 in supersymmetric models \cite{SUSYFandG}. 
In our case, both functions are reduced to 
 $\sim -\frac{1}{2 \sqrt{x}}$ for $ x \gg 1$, 
 and $\varepsilon$ can be simplified as 
\begin{equation}
\varepsilon 
\simeq \frac{3}{16\pi}\frac{1}{(hh^{\dagger})_{11}}\sum_{i=2,3}
\textrm{Im}(hh^{\dagger})^2_{1i} \frac{M_1}{M_i} .
\end{equation}
Through the relations of the seesaw mechanism, 
 this formula can be roughly estimated as 
\begin{eqnarray}
\varepsilon 
 \simeq \frac{3}{16\pi}\frac{M_1 m_3}{v^2} \sin\delta 
 \simeq 10^{-6}\left(\frac{M_1}{10^{10}\textrm{GeV}}\right)
 \left(\frac{m_3}{0.05 \textrm{eV}}\right) \sin\delta,  
 \label{epsilon}
\end{eqnarray}
where $m_3$ is the heaviest light neutrino mass, $v= 174$ GeV 
 is the vacuum expectation value (VEV) of Higgs and 
 $\sin\delta$ is an effective CP phase.  
Here we have normalized $m_3$ by $0.05$ eV 
 which is a preferable value 
 in recent atmospheric neutrino oscillation data 
 $\sqrt{\Delta m_{\oplus}^2}\simeq 0.05$ eV \cite{atm}. 
Using the above $\epsilon$, 
 the resultant baryon asymmetry generated via thermal leptogenesis 
 is described as  
\begin{equation}
\frac{n_b}{s} \simeq \frac{\varepsilon}{g_*} d  , 
 \label{b-sRatio}
\end{equation}
where $g_* \sim 100$ is the effective degrees of freedom 
 in the universe at $T \sim M_1$, 
 and $ d \leq 1 $ is the so-called dilution factor. 
This factor parameterizes how the naively expected value 
 $n_b/s \simeq \epsilon/g_*$ is reduced due to washing-out processes. 
To evaluate the resultant baryon asymmetry precisely,   
 numerical calculations are necessary, and 
 the lower bound on the lightest heavy neutrino mass 
 in order to obtain the realistic baryon asymmetry 
 in the present universe $n_b/s \simeq 10^{-10}$ 
 for fixed $\varepsilon/g_* \sin \delta \simeq 10^{-10}$ 
 has been found to be $M_1 \gtrsim 10^9$ GeV \cite{LowerBound}. 

For successful thermal leptogenesis, 
 the reheating temperature after inflation 
 should be higher than the lightest heavy neutrino mass. 
This fact causes a problem, when we consider 
 the thermal leptogenesis in supersymmetric models. 
In supersymmetric models, 
 if gravitinos produced from thermal plasma 
 decay after big bang nucleosynthesis (BBN), 
 high energy particles originating from the gravitino decay 
 would destroy light nuclei successfully synthesized by the BBN. 
In order to maintain the success of the BBN, 
 the number density of the produced gravitino 
 is severely constrained to be small. 
The resultant number density of the produced gravitino 
 is proportional to the reheating temperature, 
 and then the upper bound on the reheating temperature 
 has been found to be $T_R \lesssim 10^{6-7}$ GeV \cite{GravitinoProblem} 
 for the gravitino mass being around the electroweak scale. 
The reheating temperature should be far below 
 the lightest heavy neutrino mass. 
Therefore, in order to realize the successful thermal leptogenesis 
 in supersymmetric models, new ideas are necessary, 
 such as the gravitino 
 as the lightest supersymmetric particle (LSP) \cite{LeptoGravino}. 

In the standard cosmology, 
 the expansion of the universe is governed by
\begin{equation}
H^2= \frac{8\pi G_N}{3}\rho,
\end{equation}
The condition of the out-of-equilibrium decay 
 in Eq.~(\ref{OutOfEquilibDecay}) 
 to provide sufficient lepton asymmetry 
 without dilution $d \sim 1$ is rewritten as
\begin{eqnarray}
\tilde{m}_1 \equiv \sum_j(h_{1j}h_{1j}^\dagger)\frac{v^2}{M_1} < 
\frac{8\pi v^2}{M_1^2} \left.H\right|_{T=M_1} 
\equiv m_* \simeq 1\times 10^{-3}\textrm{eV} , \label{m*Cond}
\end{eqnarray}
 with the decay width of the lightest heavy neutrinos 
\begin{eqnarray}
 \Gamma_{N_1} = \sum_j \frac{h_{1j} h^\dagger_{1j}}{8 \pi} M_1 .
\end{eqnarray}
This condition can be regarded as the upper bound on
 the lightest neutrino mass $m_1$, 
 since the inequality $m_1 \leq \tilde{m}_1$ can be shown \cite{Davidson, Fujii}.
Considering Eq.~(\ref{epsilon}), 
 this upper bound is normally interpreted as an implication 
 that thermal leptogenesis cannot generate sufficient baryon asymmetry 
 in the case of the degenerate mass spectrum of light neutrinos 
 \cite{Fujii, DegImp}.

\section{Leptogenesis in brane world cosmology}

Let us consider the case where the lightest heavy neutrinos 
 decays in the brane world cosmology era, 
 namely $M_1 > T_t$. 
In the era, the expansion law of the universe is nonstandard 
 such as 
\begin{eqnarray} 
 H = H_{st} \sqrt{1+ \left( \frac{T}{T_t} \right)^4} 
   \simeq H_{st} \left( \frac{T}{T_t} \right)^2 .    
\end{eqnarray} 
Accordingly, 
 the condition for the out-of-equilibrium decay 
 of the heavy neutrino is modified as 
\footnote{
It is nontrivial whether the same condition of 
the out-of-equilibrium decay is applicable 
for the brane world cosmology. 
After this work is finished, 
a related preprint has appeared~\cite{Bentoetal}, 
where the same subject is addressed and 
Boltzmann equations are numerically solved. 
Their results justify our rough estimation here 
in the same manner as the one in the standard cosmology, 
though there are some discrepancies between resultant 
numerical values given by their numerical calculations 
and the ones we estimate in this paper. }
\begin{equation}
\sum_j\frac{h_{1j}h_{1j}^\dagger}{8\pi}M_1 < 
 H(T=M_1) \simeq H_{st}(T=M_1) 
 \left( \frac{M_1}{T_t} \right)^2 .  
\label{OutOEquilDecayBrane}
\end{equation}
Now we obtain the upper bound on the lightest neutrino mass 
 in the brane world cosmology such that
\begin{eqnarray}
\tilde{m}_1 < m_* \left(\frac{M_1}{T_t}\right)^2 
 \simeq 1\times 10^{-3}\textrm{eV} 
 \times \left(\frac{M_1}{T_t}\right)^2.
\label{m*NewCond}
\end{eqnarray}
Note that the upper bound has been moderated 
 due to the enhancement factor $ (M_1/T_t)^2 $ for $M_1 >T_t$. 
This result implies that, if $T_t$ is low enough, 
 the thermal leptogenesis scenario is successful 
 even in the case of the degenerate light neutrino mass spectrum. 

In the above discussion, we have implicitly assumed that 
 the lightest heavy neutrino is in thermal equilibrium 
 at a high temperature $T > M_1$. 
In the following, let us verify whether this situation 
 can be realized in the brane world cosmology. 
Since the right-handed neutrinos are singlet 
 under the SM gauge group, 
 the only interaction through which the right-handed neutrino 
 can be in thermal equilibrium is the Yukawa coupling 
 in Eq.~(\ref{Yukawa}).
\footnote{
If we extend our model into a model 
 such as the left-right symmetric model, 
 we can consider the case where 
 the right-handed neutrinos can be in thermal equilibrium  
 through new gauge interactions \cite{Plumacher}.
} 
Consider a pair annihilation process of 
 the lightest heavy neutrino through the Yukawa couplings. 
The thermal averaged annihilation rate is roughly estimated as 
\begin{eqnarray}
\Gamma_{int}(T) = n \langle\sigma v\rangle  
\simeq 
\frac{3 \zeta(3)}{2 \pi^2} T^3 
 \times 
\frac{N g_Y^4}{100} T^{-2} ,  
\end{eqnarray}
 where $N$ is the number of annihilation channels, 
 $g_Y $ stands for dominant Yukawa couplings, 
 and the factor $1/100$ denotes the kinematical phase factor. 
The freeze-in temperature, $T_{FI}$, 
 in the brane world cosmology era 
 can be defined from the condition 
 $\Gamma_{int}(T_{FI}) = H (T_{FI}) 
 \simeq H_{st}(T_{FI}) (T_{FI}/T_t)^2$, 
 and we obtain 
\begin{equation}
T_{FI}  \simeq 1.1 \times 10^{10} \mbox{GeV} 
 \times 
 \left( \frac{N g_Y^4}{10} \right)^{1/3}   
 \left( \frac{g_*}{100} \right)^{-1/6}   
 \left( \frac{T_t}{10^7 \mbox{GeV}} \right)^{2/3}.   
\label{FItemp}
\end{equation}
Here we have used a large value 
 for the normalization of the Yukawa coupling constant. 
It is necessary to fine-tune each elements of large Yukawa couplings, 
 in order to obtain masses of the light neutrinos much smaller 
 than the scale naively obtained through the see-saw mechanism. 
Large Yukawa couplings might be reasonable 
 because the thermal leptogenesis with the degenerate 
 light neutrino mass spectrum is possible 
 in the brane world cosmology as discussed above. 
Consistency of our discussion, namely $T_t \leq T_{FI}$, 
 leads to the upper bound on the transition temperature 
 such as 
\begin{equation}
T_t \lesssim 1.3 \times 10^{16} \mbox{GeV} 
 \times 
 \left( \frac{N g_Y^4}{10} \right) 
 \left( \frac{g_*}{100} \right)^{-1/2} . 
\label{Ttmax}
\end{equation} 
The thermal leptogenesis in the brane cosmology era 
 can take place if the lightest heavy neutrino mass is in the range 
\begin{equation}
    T_t < M_1 < T_{FI} .  
\label{MassRange}
\end{equation}
Recall that $M_1 \simeq 10^{9-10}$ GeV is required 
 in order to generate the sufficient baryon asymmetry. 
This implies $10^{9-10}$ GeV $> T_t \gtrsim 10^{6-7}$ GeV 
 ($10^{12-13}$ GeV $> M_5 \gtrsim 10^{10-11}$ GeV ) from Eqs.~(\ref{FItemp}) and (\ref{MassRange}). 

Finally, there is an additional interesting possibility 
 for the thermal leptogenesis in the supersymmetric case. 
In the standard cosmology, 
 the thermal leptogenesis in supersymmetric models 
 is hard to be successful, 
 since the reheating temperature after inflation 
 is severely constrained to be $T_R \lesssim 10^{6-7}$ GeV 
 due to the gravitino problem \cite{GravitinoProblem}. 
However, as pointed out in Ref.~\cite{OSGravitino}, 
 the constraint on the reheating temperature 
 is replaced with the one on the transition temperature 
 in the brane world cosmology. 
Therefore, 
 when the transition temperature is low enough 
 $T_t \lesssim 10^{6-7}$ GeV, 
 the gravitino problem can be solved 
 even if the reheating temperature is much higher. 
Interestingly, the transition temperature 
 $T_t \simeq 10^{6-7}$ GeV we found above 
 can realize the thermal leptogenesis scenario successfully 
 and also solve the gravitino problem. 
However, for this scenario, one should notice that inflation need to have 
 a very efficient reheating with $T_R \simeq 10^{9-10}$ GeV, 
 because the potential energy during inflation $V_{inf}$ 
 must be less than $M_5^4 \simeq (10^{10-11}$ GeV $)^4$ 
 in order for inflation models to be consistently treated 
 in the context of the five-dimensional theory. 
In fact, such inflation models are possible but limited \cite{CS}.
Further studies on inflation would provide informations for this possibility.

\section{Conclusion} 

We have studied the thermal leptogenesis in the brane world cosmology. 
The nonstandard expansion law of the brane world cosmology
 affects the condition of the out-of-equilibrium decay 
 of the lightest heavy neutrino, 
 and moderates the upper bound on the lightest light neutrino mass. 
As a result, the degenerate mass spectrum for the light neutrinos 
 can be consistent with the successful thermal leptogenesis scenario, 
 if the transition temperature is lower than 
 the lightest heavy neutrino mass. 
We have verified that 
 the lightest heavy neutrino can be in thermal equilibrium 
 through the Yukawa couplings among light neutrinos and Higgs bosons
 and found the region of the transition temperature 
 consistent with the successful thermal leptogenesis 
 in the brane world era. 
Then, we obtain the constraint on $T_t$ ($M_5$). 
Furthermore, in supersymmetric case, 
 we have noticed that the transition temperature 
 required for the successful thermal leptogenesis 
 can solve the gravitino problem simultaneously.

%
\section*{Acknowledgments}
The work of N.O. is supported in part by the Grant-in-Aid 
 for Scientific Research  in Japan (\#15740164). 
The work of O.S. is supported by PPARC. 
The authors thank the Yukawa Institute for Theoretical Physics 
 at Kyoto University, where this work was completed 
 during the YITP Workshop (YITP-W-05-02) 
 on "Progress in Particle Physics 2005".



\end{document}